\documentclass{article}

\usepackage[utf8]{inputenc} 
\usepackage{geometry} 
\usepackage{graphicx} 
\usepackage{amsmath} 
\usepackage{hyperref} 
\usepackage{verbatim} 
\usepackage{listings} 
\usepackage{xcolor} 
\usepackage{changepage}
\usepackage{tabularx}
\usepackage{booktabs}
\usepackage{float}
\usepackage{array} 
\newcolumntype{C}{>{\centering\arraybackslash}X}
\newcolumntype{Y}{>{\centering\arraybackslash}X}
\title{A Narrative Review of Identity, Data and Location Privacy Techniques in Edge Computing and Mobile Crowdsourcing}
\author{
  Syed Raza Bashir \\
  Department of Computer Science, \\
  Toronto Metropolitan University, \\
  \texttt{syedraza.bashir@torontomu.ca}
  \and
  Shaina Raza \\
  Department of Computer Science, \\
  Toronto Metropolitan University, \\
  \texttt{shaina.raza@torontomu.ca}
  \and
  Vojislav Misic \\
  Department of Computer Science, \\
  Toronto Metropolitan University, \\
  \texttt{vmisic@torontomu.ca}
}
\date{\today}

\begin{document}

\maketitle
\begin{abstract}
    
As digital technology advances, the proliferation of connected devices poses significant challenges and opportunities in mobile crowdsourcing and edge computing. This narrative review focuses on the need for privacy protection in these fields, emphasizing the increasing importance of data security in a data-driven world. Through an analysis of contemporary academic literature, this review provides an understanding of the current trends and privacy concerns in mobile crowdsourcing and edge computing. We present insights and highlight advancements in privacy-preserving techniques, addressing identity, data, and location privacy. This review also discusses the potential directions that can be useful resources for researchers, industry professionals, and policymakers.
\end{abstract}

\section{Introduction}
The advent of the digital era and information overload has resulted in a significant increase in connected devices, generating a substantial volume of data transmitted over the Internet~\cite{bashir2023bert4loc}. Managing predominantly through cloud infrastructures, the ubiquity of smart devices such as smartphones, tablets, smartwatches, and fitness trackers has made it even more complex. These devices routinely gather extensive contextual information about users, including their locations, activities, and environmental conditions—data that are vital for applications aimed at predicting user behavior and delivering personalized experiences~\cite{adomavicius2010context,raza2019progress}.

In response to these developments, mobile crowdsourcing has emerged as a pivotal solution. This approach involves individuals collectively contributing data via various digital platforms~\cite{liu2019unitask}. Applications such as traffic monitoring systems leverage crowdsourced data to provide real-time insights. However, these methods also raise significant concerns regarding data privacy and the risks of unauthorized access to sensitive information~\cite{bashir2022improving, hu2010privacy}.

The rapid increase in mobile data traffic poses substantial challenges for data management and processing. Edge computing has been identified as a promising solution to these challenges by processing data closer to its source rather than relying solely on central servers. This approach helps alleviate network congestion but also introduces new privacy and security vulnerabilities that must be addressed~\cite{zhang2018data}.

This paper explores the intersection of mobile crowdsourcing and edge computing with a particular emphasis on privacy preservation. It highlights the potential for integrating these technologies to enhance data processing efficiency while maintaining user privacy. We provide an overview of existing research on these technologies and their implications for privacy. Our goal is to equip researchers, practitioners, and policymakers with the necessary insights to navigate the complexities of ensuring privacy while advancing technological capabilities. Additionally, we discuss potential strategies for mitigating privacy risks associated with decentralized data processing and propose future research directions to address these challenges comprehensively.

\subsection{Literature Review Strategy}
\begin{figure}[h]

        \includegraphics[width=.98\linewidth]{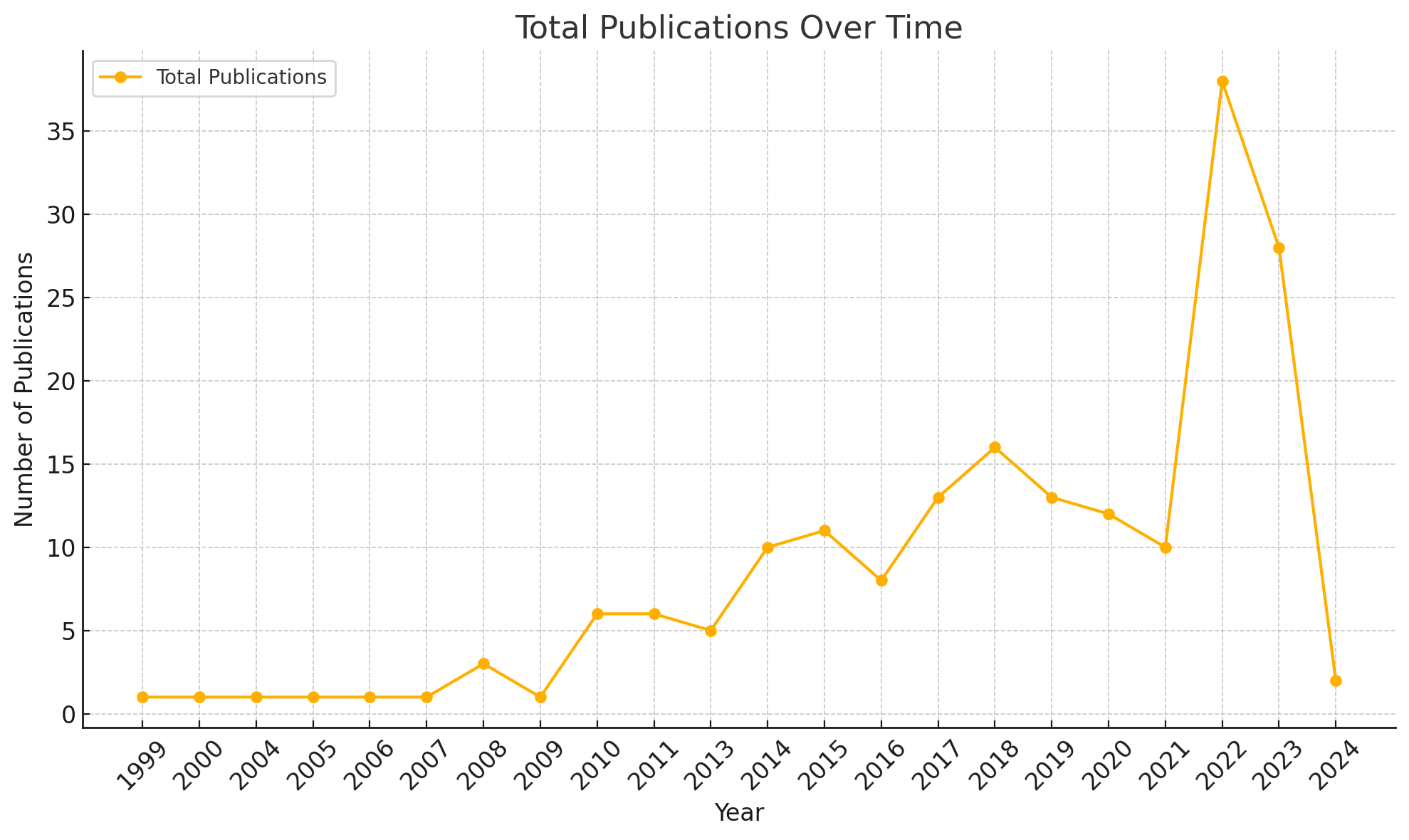}
        \caption{Publication trend on privacy in edge computing and mobile crowdsourcing in recent years. The full year 2024 is not covered, so additional publications are expected.}
        \label{fig:bib-a1}

\end{figure}

\begin{figure}[h]
   
        \includegraphics[width=.98\linewidth]{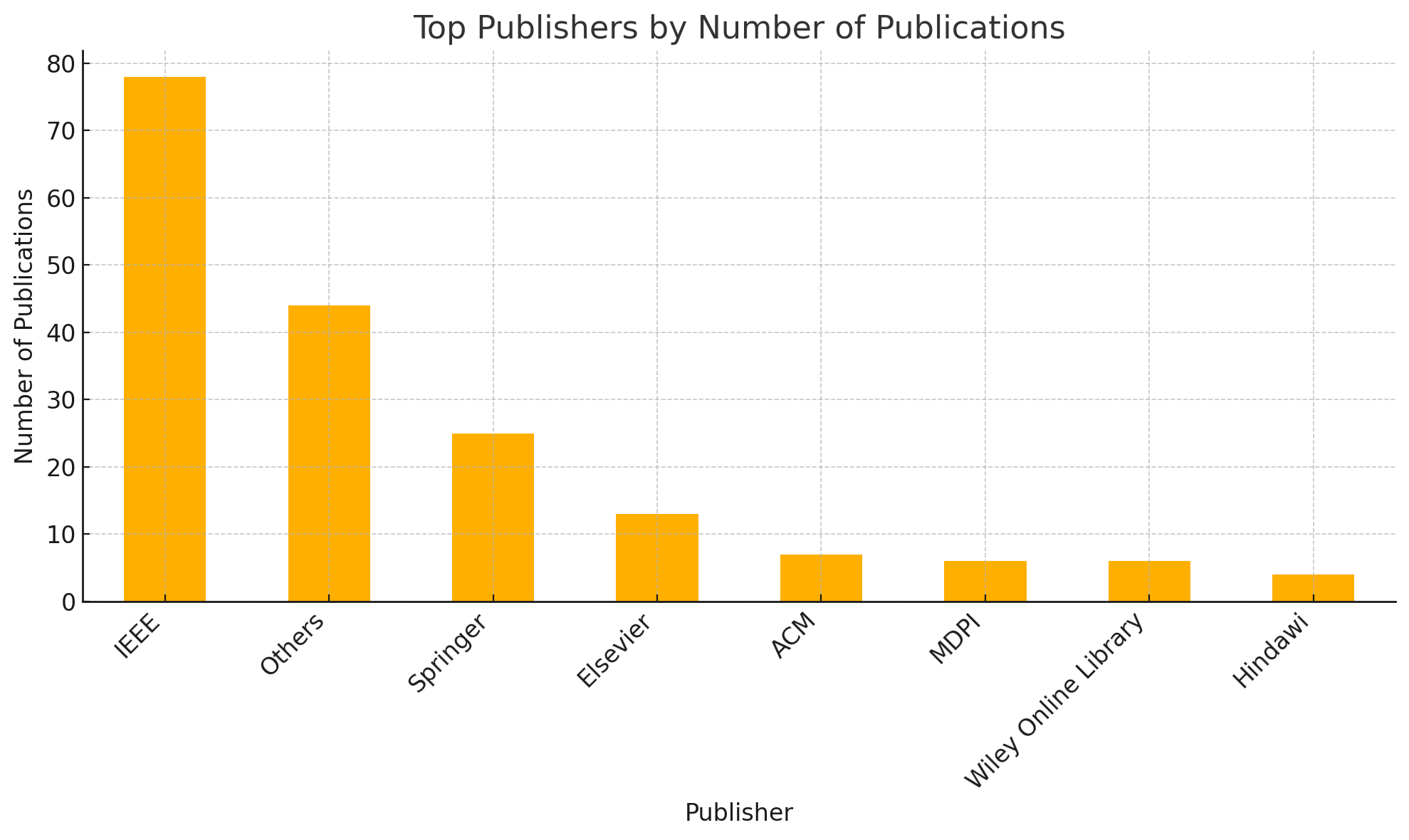}
        \caption{Top publishers by number of publications in the field.}
        \label{fig:bib-a3}

\end{figure}

\begin{figure}[h]
        \includegraphics[width=.98\linewidth]{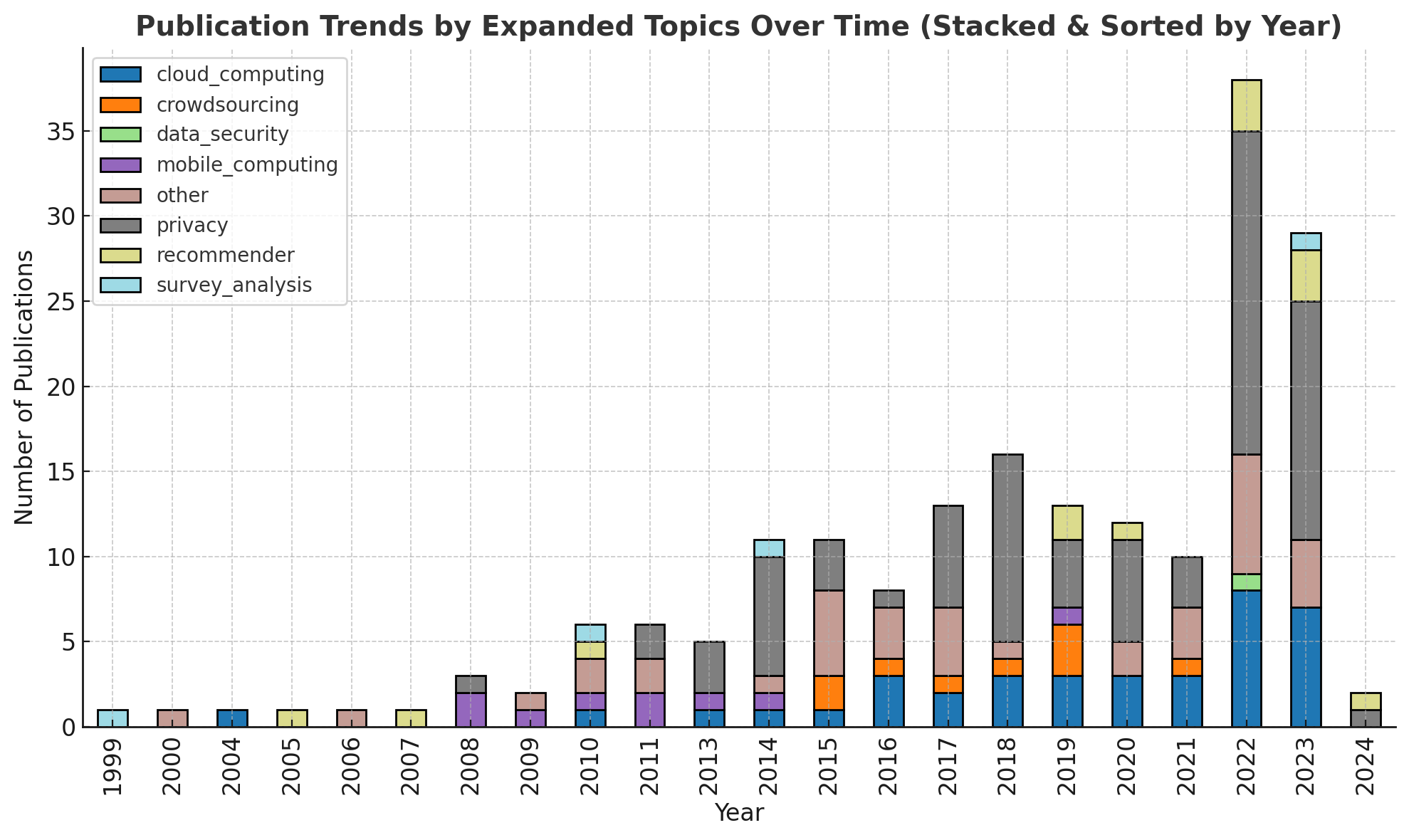}
        \caption{Publication trends over time by topic. }
        \label{fig:bib-a2} 
\end{figure}

Our literature review strategy involved searches in academic repositories such as ACM Digital Library, IEEE Xplore, Springer, Elsevier, Google Scholar, DBLP, CiteSeerX, Microsoft Academic Search, Web of Science, and ScienceDirect. Publications from the period 2012 to 2021 were primarily considered, with a final selection of approximately 52 peer-reviewed papers from an initial pool of 150 papers based on relevance to mobile crowdsourcing and edge computing. These papers were analyzed for trends, methodologies, findings, and gaps in research. The publication trend in the last few years is given in Figures~\ref{fig:bib-a1}--\ref{fig:bib-a2}.
\vspace{-12pt}

We used queries to search for relevant literature , shown in Listing \ref{lst:search-queries}:
\begin{lstlisting}[caption={Search Query Strings for Mobile Crowdsourcing and Edge Computing Studies}, label={lst:search-queries}]
1. ("Mobile Crowdsourcing" AND "Edge Computing" AND "Privacy Preservation") AND ("Data Security" OR "Real-Time Data Processing") AND (2012-2023)

2. ("Mobile Crowdsourcing" AND "Edge Computing") AND ("User Context" OR "Personalized Services") AND ("IoT" OR "Cloud Computing") AND (peer-reviewed)

3. ("Mobile Crowdsourcing" AND "Edge Computing") AND ("Privacy" AND "Information Overload") AND ("Smart Devices" OR "User Behavior Prediction") AND (2012-2023)
\end{lstlisting}

\subsection{Contributions}
This work contributes the following to the field:
\begin{enumerate}
    \item   This review offers an overview of mobile crowdsourcing and edge computing, with an emphasis on the importance of privacy preservation.
    \item We synthesize both contemporary and foundational literature to provide an understanding of identity privacy, data privacy, and location privacy (based on their relevance to importance) and identify current research trends.
\end{enumerate}
A meta-review of previous surveys is presented in Table~\ref{tab:lit}, and the main keywords used in this work are given in Table~\ref{app:definitions}.

\begin{table}[H]
\caption{Related surveys.}
\centering
\begin{tabularx}{\textwidth}{|Y|Y|Y|}
\toprule
\textbf{Domain} & \textbf{Focus} & \textbf{Key Contributions and Privacy Aspects} \\
\midrule
\textbf{Edge Computing} & Real-time and local data processing, IoT integration, Edge AI, Security & Studies highlighted capabilities for local processing, integration with AI, and identified gaps in privacy protection specific to edge computing environments. Key studies include~\cite{10169821,10026418,EdgeAI2023,mukherjee2017security,guan2018data,zhang2018data,yousefpour2019all,zeyu2020survey}. \\
\midrule
\textbf{Mobile Computing} & Security Trends, Healthcare, Cloud Integration, AR Applications & Focused on evolving security challenges, particularly in mobile cloud contexts and healthcare applications during the COVID-19 pandemic, and AR enhancements. Relevant studies include~\cite{engproc2023032022,ALI2023109605,jararweh2016future,irshad2017advances,liu2021cloud}. \\
\bottomrule
\end{tabularx}
\label{tab:lit}
\end{table}

\subsection{Necessity of This Review}
\begin{figure}[h]
     
    \includegraphics[width=.98\linewidth]{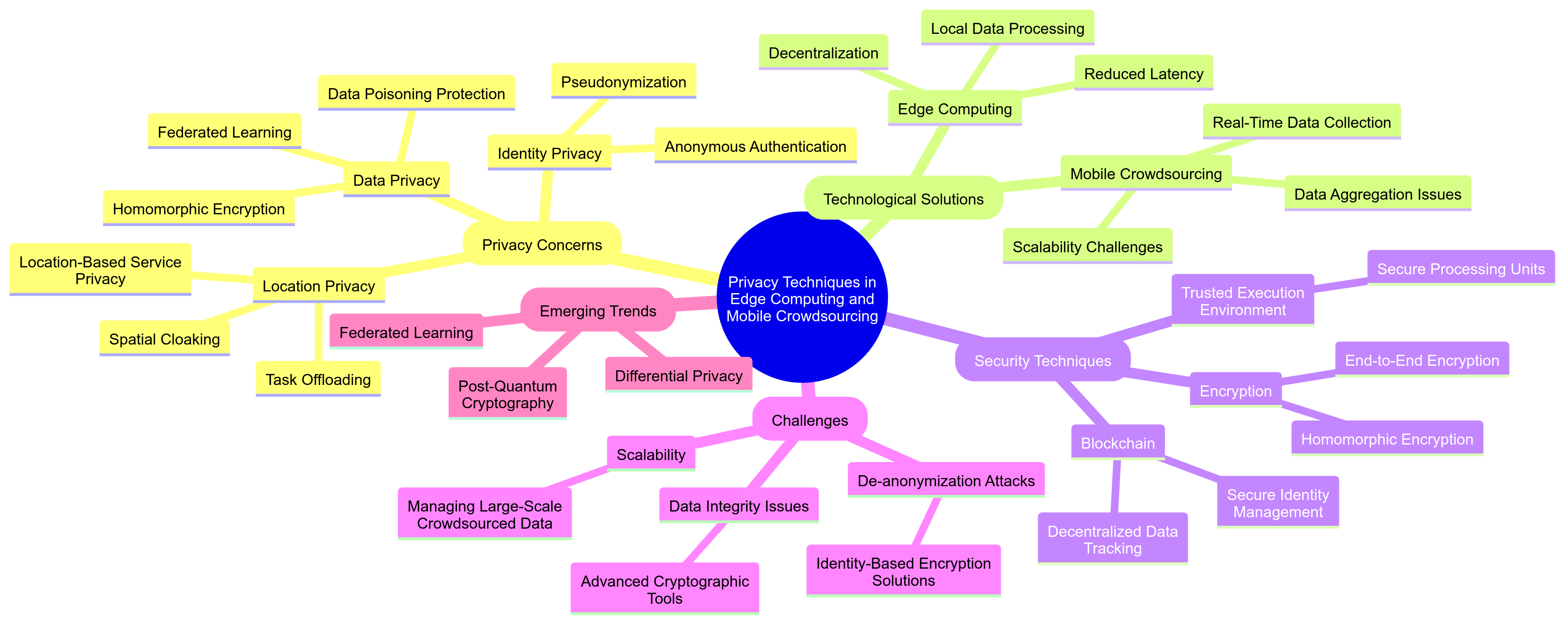}
    \caption{Flow of the privacy-preserving methods covered in this work.}
    \label{fig:main}
\end{figure}

This review addresses a literature gap that surrounds the rapidly evolving fields of mobile crowdsourcing and edge computing. As our reliance on data-driven systems continues to grow, this review provides a timely analysis of current trends, privacy concerns, and research gaps in these interconnected domains. We also shed some light on key privacy issues, including those related to identity, data, and location, and offer some information to researchers, industry professionals, and policymakers. 
The flow of this review is shown in Figure~\ref{fig:main}.
 
\section{Privacy Concerns in Mobile Crowdsourcing and Edge Computing}
We discuss mobile crowdsourcing and edge computing and examine privacy challenges in these domains, focusing on identity privacy, data privacy, and location privacy due to their significant impact on protecting users in these interconnected systems. 

\subsection{Identity Privacy}
Identity privacy is crucial in protecting individual personal information from being exposed or inferred during mobile crowdsourcing or edge computing activities~\cite{yu2014identity}. This protection encompasses safeguarding personal details such as names, social security numbers, and other identifiers that could link activities to a specific individual.

In mobile crowdsourcing, data are often collected from numerous participants, making it essential to ensure their anonymity or pseudonymity. This is critical to prevent the misuse of personal information~\cite{veeresh2023identity}. If personal identifiers are exposed, it can lead to identity theft, where malicious actors might use stolen identities to commit fraud or other illegal activities~\cite{suresha2022ensuring}. Thus, safeguarding identity privacy is fundamental for establishing and maintaining user trust.

The complexity of data collection processes in mobile crowdsourcing also raises concerns about implied consent and the need for more transparent consent management. Often, users agree to data collection practices without fully understanding the implications, inadvertently compromising their privacy. Therefore, it is vital to develop clearer consent mechanisms that ensure users are informed and their privacy is respected throughout the data collection process.

\subsection{Data Privacy}
Data privacy in the context of mobile crowdsourcing and edge computing involves protecting the data collected, processed, and stored from unauthorized access, alteration, or disclosure~\cite{wang2022dpp}. This protection is crucial, as crowdsourced data often include sensitive information such as health records, financial transactions, and personal communications. Ensuring the confidentiality and integrity of such data is vital to prevent unauthorized access and maintain user trust~\cite{wang2022dpp}.

Various regulations, such as the General Data Protection Regulation (GDPR) (\url{https://gdpr-info.eu/}) (accessed on 22\textsuperscript{nd} October, 2024) 
 and the Health Insurance Portability and Accountability Act (HIPAA) (\url{https://www.hhs.gov/hipaa/}) (accessed on 22\textsuperscript{nd} October, 2024), 
 mandate strict measures to safeguard data privacy. Breaches can lead to severe consequences, including financial losses, reputational damage, and erosion of user trust. A significant concern in this domain is the risk associated with third-party data sharing. Once data enter the crowdsourcing ecosystem, users often lose control over their personal information as it flows through multiple parties in edge computing environments. This lack of control raises privacy risks that need to be addressed.

Furthermore, mobile crowdsourcing systems are vulnerable to specific attacks like Sybil attacks and data poisoning~\cite{kumar2021blockchain}. These attacks can manipulate the crowdsourcing process and potentially expose sensitive user information. Addressing these vulnerabilities requires robust security measures to protect against unauthorized data access and ensure data integrity throughout the crowdsourcing lifecycle.

\subsection{Location Privacy}
Location privacy is the protection of information regarding an individual geographical location, particularly as it is obtained through mobile crowdsourcing and edge computing. This involves ensuring that location data cannot be used to track or monitor an individual's movements without their explicit consent. Unauthorized access to location data poses significant risks, including stalking, harassment, and other malicious activities where an individual's movements might be tracked without their knowledge or consent~\cite{liu2022privacy}. The disclosure of sensitive locations, such as homes, workplaces, or places of interest, can lead to security risks and privacy violations~\cite{zhang2022location}. Therefore, it is crucial for users to have control over who can access their location data and for what purposes. Ensuring location privacy is fundamental to preserving users' autonomy and their right to control their personal information~\cite{ko2020lpga}.

The risks associated with sharing location data are many, for example, malicious actors can use precise location information to deduce behavioral patterns and predict where someone might be at any given time, potentially facilitating physical tracking or blackmail~\cite{liu2022privacy}. Additionally, unauthorized sharing or breaches of location data can lead to unintended surveillance and discrimination by authorities or third parties~\cite{yang2024ai}.

To mitigate these risks, it is essential for organizations to implement robust privacy measures. These include obtaining explicit user consent before enabling location-tracking features and ensuring transparency about how location data are used and shared. Techniques such as spatial cloaking and cryptographic methods can also enhance location privacy by anonymizing data and securing communication channels~\cite{gambs2014anonymization}.

\subsubsection*{Adhoc Solutions} 
To address privacy concerns in mobile crowdsourcing and edge computing, researchers and practitioners are developing several innovative solutions.

One approach is the implementation of decentralized systems~\cite{cutillo2009privacy}, which enhance user control over data and minimize the risks associated with centralized data repositories. This helps prevent vulnerabilities that arise from having a single point of failure or attack. Organizations are also shifting towards using advanced encryption techniques~\cite{brakerski2011fully} to protect data both when it is stored and during transfer. The goal is to ensure that even if someone intercepts the data, it remains unintelligible without proper authorization, thereby safeguarding sensitive information.
Technologies like differential privacy~\cite{jiang2021privacy} help in protecting sensitive information while still allowing for meaningful data analysis. By adding noise to datasets, these techniques make it difficult to identify individual users, thus preserving privacy while enabling valuable insights. 

There is an increasing focus on creating clear and easy-to-use systems for managing consent~\cite{zhao2022crowdfl}. These frameworks ensure that users are fully informed about data-sharing practices and can easily control their preferences. This empowers users to have more say over their personal information and its handling, especially on crowdsourcing platforms. 
Also, leveraging blockchain technology~\cite{kumar2021blockchain} provides traceability and immutability in data handling practices. Blockchain enhances security and transparency, allowing users to verify how their data are used and shared. This not only improves security but also builds trust by providing a transparent record of transactions and data usage.

\section{Mobile Crowdsourcing}

Mobile crowdsourcing represents a field where collective user contributions, primarily via mobile devices, drive data collection and service provision~\cite{guo2015mobile, ren2015exploiting}. Mobile crowdsourcing leverages the ubiquitous nature of mobile devices into a vast pool of user-submitted data, encompassing geospatial and behavioral data as well as environmental and health metrics. Its applications are vast, covering environmental monitoring, disaster response, urban planning, healthcare, and transportation, and highlight the versatility and impact of mobile crowdsourcing~\cite{kong2019mobile}.

A quintessential example of the potential of mobile crowdsourcing is seen in Waze~\cite{waze}, a navigation app that integrates user-generated traffic data to provide real-time routing advice and traffic condition updates. This not only showcases the practical implementation of mobile crowdsourcing but also demonstrates how user participation can significantly enhance service quality and responsiveness.

\begin{figure}[h]
     
    \includegraphics[width=.98\linewidth]{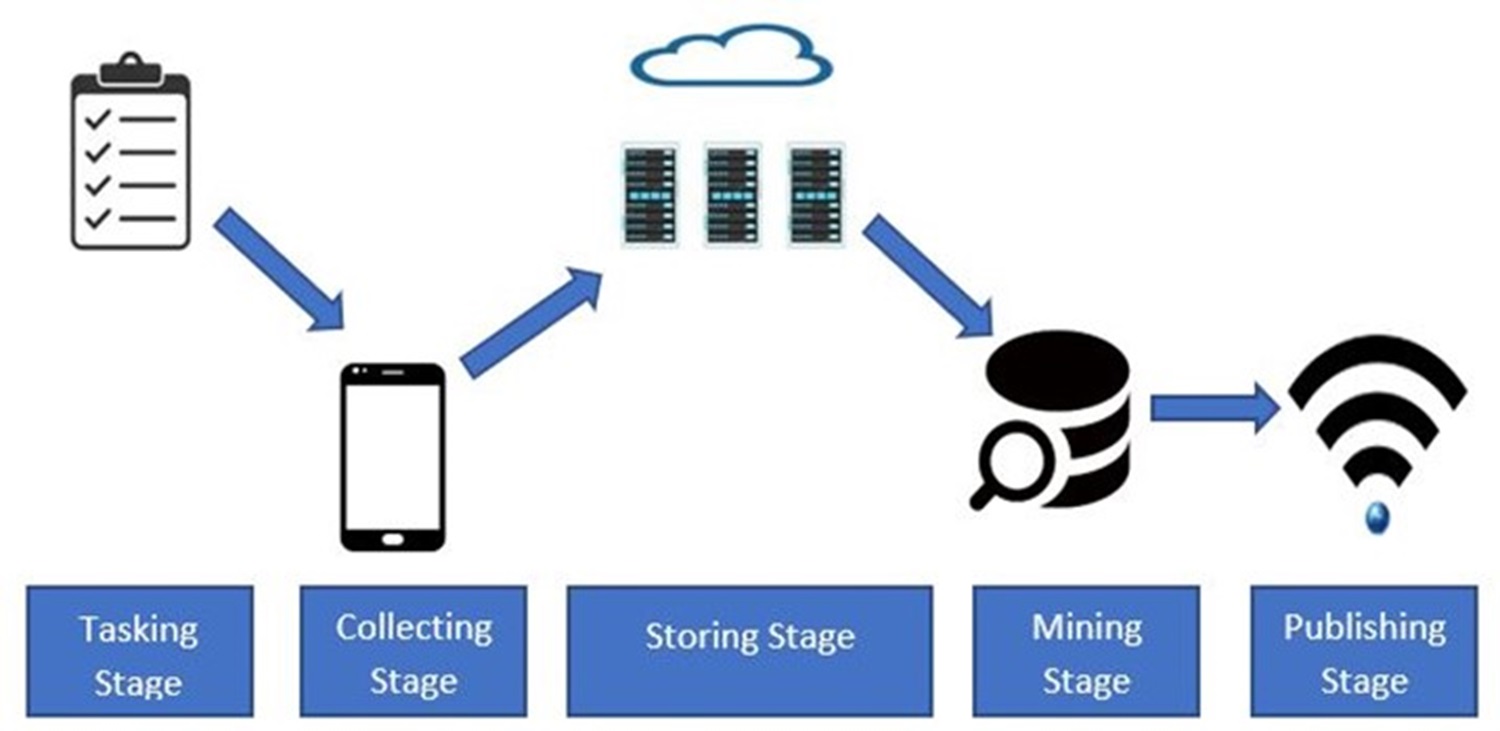}
    \caption{Anatomy of a mobile crowdsourcing campaign.}
    \label{fig:mobile-crowdsourcing-anatomy}
\end{figure}

We show the key stages in mobile crowdsourcing in Figure~\ref{fig:mobile-crowdsourcing-anatomy}.
\begin{enumerate} 
    \item  Tasking Stage: This initial phase involves defining clear and actionable data collection objectives that align with the overarching goals of crowdsourcing.
    \item Collecting Stage: At this point, the data are actively acquired from users through their mobile devices. The data can range from environmental readings to user-generated content and location data, providing a rich dataset for analysis.
    \item Storing Stage: The collected data are centralized on cloud or edge servers for processing. This stage is crucial for managing the volume of data and ensuring it is accessible for~analysis.
    \item Mining Stage: Advanced analytics, machine learning (ML), and data mining techniques are applied to the aggregated data to extract meaningful insights, patterns, and trends.
    \item Publishing Stage: The final stage involves disseminating the findings and insights to end-users or stakeholders, often through the same mobile platforms used for data~collection.
\end{enumerate}

The technological infrastructure supporting mobile crowdsourcing includes a range of components from mobile devices and network communication protocols to cloud and edge computing platforms. These technologies facilitate the efficient collection, transmission, and processing of crowdsourced data. However, this complex ecosystem also introduces challenges related to scalability, data quality control, and the integration of heterogeneous data sources. The transmission of sensitive personal information, such as location data and personal identifiers, to centralized servers poses risks related to data security and user privacy~\cite{a2020comprehensive,abbas2017mobile}.

Literature: In mobile computing, recent surveys have focused on security trends, healthcare applications during the COVID-19 pandemic, and the integration with cloud computing. One study~\cite{engproc2023032022} explored evolving security challenges in mobile cloud computing and strategies for safeguarding these environments. The significant role of mobile computing in healthcare, especially during the pandemic, was highlighted in a survey~\cite{ALI2023109605}, addressing both challenges and potential solutions. Additionally, the integration of mobile computing with edge computing is examined in~\cite{jararweh2016future}, where the authors discuss the benefits of leveraging edge computing for reducing latency and improving processing capabilities in mobile applications. The survey~\cite{irshad2017advances} delves into the advancements in mobile augmented reality (AR), focusing on the integration of mobile computing with AR to enhance user experiences and application performance. Further research~\cite{liu2021cloud} reviews the application of mobile computing in smart cities.


\subsection{Privacy Concerns in Mobile Crowdsourcing}

\subsubsection{Identity Privacy in Mobile Crowdsourcing}
Identity privacy in mobile crowdsourcing is an important aspect of safeguarding user data, which includes sensitive information like names, addresses, and contact details. Discussions on identity privacy focus on the implementation of robust forward security measures to protect cloud data~\cite{li2023privacy, veeresh2023identity}. Identity-based encryption methods are highlighted as a means to secure cloud data, which shows the multifaceted approach required to enhance privacy protection in cloud computing contexts~\cite{suresha2022ensuring}. Consolidated Identity Management Systems (CIDM) are proposed to safeguard user identity by separating authorization credentials, adding an extra authentication layer, and securing communication links with cloud service providers~\cite{khalil2014consolidated}. Further discussions emphasize blockchain-based trust models and strategies for identity management that negate the need for trusted third parties~\cite{khajehei2018preserving, bendiab2018wip}.

\subsubsection{Data Privacy in Mobile Crowdsourcing}
Data privacy in mobile crowdsourcing involves developing methods and techniques focused on enhancing data security and privacy within cloud environments. Research efforts include sophisticated methods for conducting spatial data queries that prioritize user privacy without compromising data utility~\cite{miao2023efficient}. Techniques leveraging advanced cryptographic tools such as homomorphic encryption show the potential for employing encryption to fortify privacy measures in cloud computing~\cite{wang2022dpp}.

\subsubsection{Location Privacy in Mobile Crowdsourcing}
Ensuring the privacy of location data in mobile cloud computing involves various innovative methods such as cipher-text retrieval techniques for location privacy protection~\cite{zhang2022location}. Other approaches include the use of spatial cloaking to anonymize location data for queries in cloud environments~\cite{jadallah2019spatial}. Advanced cryptographic techniques, such as SHA-2 for robust mutual authentication and location privacy, are employed to ensure secure communication and privacy preservation in mobile cloud computing settings~\cite{rana2020mutual}.

Collectively, these research efforts illustrate the diverse strategies being explored to protect identity, data, and location privacy in cloud computing. These methods are summarized in Table~\ref{tab:privacy_cloud_computing}.

\begin{table}[H]

\caption{Privacy concerns and solutions in mobile cloud computing.}


\begin{tabularx}{\textwidth}{|C|C|C|C}
\toprule
 
\textbf{Privacy Concern} & \textbf{Research Focus} & \textbf{Key Solutions} \\
\midrule
Identity Privacy in Cloud Computing& Safeguarding sensitive user data such as names, addresses, and contact details & 
\begin{itemize}\vspace{-6pt}
    \item Robust forward security measures~\cite{li2023privacy, veeresh2023identity}
    \item Identity-based encryption methods~\cite{suresha2022ensuring}
    \item Consolidated Identity Management Systems (CIDM)~\cite{khalil2014consolidated}
    \item Blockchain-based trust models and identity management~\cite{khajehei2018preserving, bendiab2018wip}\vspace{-12pt}
\end{itemize} \\
\midrule

Data Privacy in Cloud Computing& Enhancing data security and privacy within cloud environments & 
\begin{itemize}\vspace{-6pt}
    \item Spatial data query methods prioritizing user privacy~\cite{miao2023efficient}
    \item Advanced cryptographic tools such as homomorphic encryption~\cite{wang2022dpp}\vspace{-12pt}
\end{itemize} \\
\midrule
Location Privacy in Cloud Computing & Ensuring privacy of location data & 
\begin{itemize}\vspace{-6pt}
    \item Cipher-text retrieval techniques for location privacy protection~\cite{zhang2022location}
    \item Spatial cloaking to anonymize location data~\cite{jadallah2019spatial}
    \item SHA-2 for robust mutual authentication and location privacy~\cite{rana2020mutual}\vspace{-12pt}
\end{itemize} \\
\midrule
\end{tabularx}
\label{tab:privacy_cloud_computing}
\end{table}

\section{Edge Computing and Privacy Concerns}

Edge computing represents a paradigm shift in data processing and management. It emphasizes the decentralization of computational resources to bring data storage and processing closer to the sources of data~\cite{zhang2018data}. This approach is particularly beneficial for real-time and latency-sensitive applications across diverse domains such as autonomous vehicles, smart homes, emergency response systems, smart agriculture, and modernized power grids.

Conceptual Framework: Edge computing seeks to address the inherent limitations of centralized cloud computing models by distributing computational tasks to the edge of the network~\cite{gao2022ppo2}. This strategic distribution significantly reduces latency by minimizing the distance between the data source and the processing unit and alleviates bandwidth constraints on the network. Such enhancements boost overall system responsiveness and~reliability.

\begin{figure}[h]
    
    \includegraphics[width=.981\linewidth]{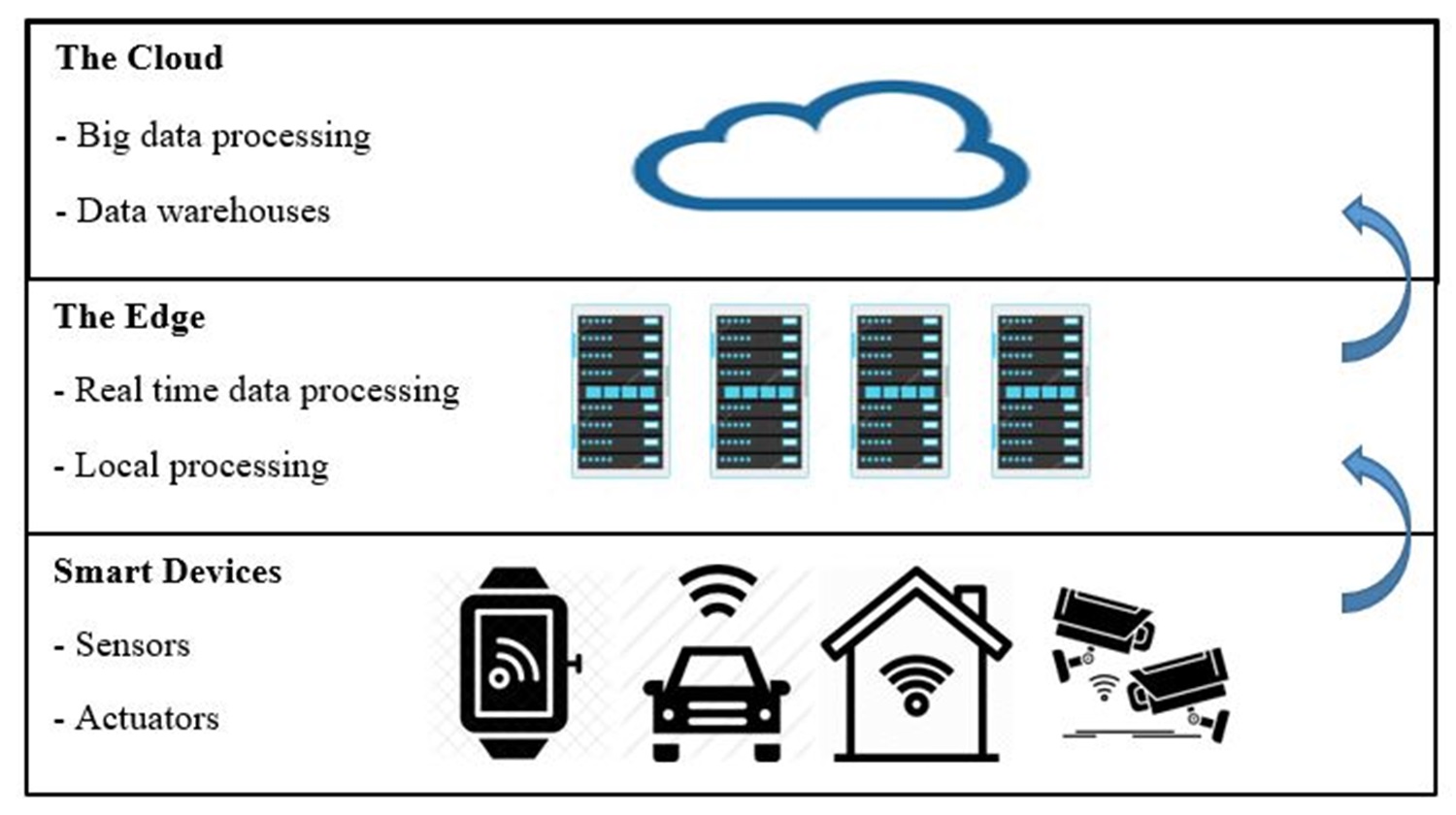}
    \caption{Edge computing architecture illustrating the interplay between user, edge, and cloud layers.}
    \label{fig:archi}
\end{figure}

Technological Advancements and Architectural Design: The architectural design of edge computing is inherently modular and scalable. As depicted in Figure~\ref{fig:archi}, the typical architecture comprises three primary layers: the user layer, edge layer, and cloud layer. Each layer plays a distinct role in the ecosystem, with smart devices at the user layer generating data, edge servers at the edge layer processing this data in real time, and cloud servers at the cloud layer providing long-term storage and further analytics capabilities.

Benefits of Edge Computing: The decentralization inherent in edge computing brings several benefits. By processing data closer to its source, it significantly reduces latency, which is crucial for applications that require real-time feedback, such as autonomous driving and emergency response systems~\cite{luo2021resource}. Furthermore, edge computing contributes to bandwidth efficiency by minimizing the volume of data that needs to be transmitted to the cloud, thereby reducing network congestion and associated costs.

Integration with Smart Technologies: Edge computing is integral to the evolution of smart technologies~\cite{liu2021cloud}. It enables the seamless integration of IoT devices, smart sensors, and other data sources, facilitating the development of intelligent applications that can operate autonomously and interact with their environment in real time. This integration is crucial for the advancement of smart cities, industrial automation, and personalized healthcare, among other applications.

Literature: Recent studies, such as those by~\cite{10169821,10026418}, distinguish edge computing from cloud computing by detailing its capabilities for real-time and local data processing, highlighting emerging research, and practical applications. Research in~\cite{EdgeAI2023} explores 'Edge AI', examining how edge computing and AI are integrated, their applications, and the impact of AI algorithms on decision-making and data processing at the network edge. Issues of security and privacy specific to edge computing are addressed in~\cite{mukherjee2017security}, where it is noted that conventional cloud computing privacy methods are inadequate. Ref.~\cite{guan2018data} highlights concerns about location privacy in edge computing environments, while~\cite{zhang2018data} reviews security and privacy preservation methods, noting a predominant focus on cloud computing over edge-specific strategies. Additionally, a survey~\cite{yousefpour2019all} discusses edge computing's influence on IoT development, advocating for its suitability over other computing paradigms for IoT scenarios. Furthermore,~\cite{zeyu2020survey} presents a comprehensive review of current research on security in edge computing, pinpointing challenges in new models, application scenarios, and technological environments, with specific attention to access control, key management, privacy protection, attack mitigation, and anomaly detection.

\subsection{Privacy Concerns in Edge Computing}
Edge computing, despite its benefits, introduces complex privacy issues due to its decentralized nature. This subsection delves into identity, data, and location privacy, highlighting ongoing research and emerging solutions aimed at enhancing security and privacy in edge environments.

\subsubsection{Identity Privacy in Edge Computing}
{Several studies have explored pseudonym technology, cryptographic methods, and data aggregation schemes in edge computing environments to preserve identity   privacy~\cite{lysyanskaya2000pseudonym, wang2017secure, lu2011pseudonym}. However, these methods are challenged by de-anonymization   attacks~\cite{gambs2014anonymization, narayanan2008robust, nilizadeh2014community} and limited data integrity~\cite{wang2017secure}. Notably, research efforts are being made to enhance security and privacy through various methods focused on maintaining anonymity, preventing unauthorized access, and ensuring user data integrity. Innovations include anonymous authentication mechanisms leveraging identity-based cryptography for secure device-to-edge server communications~\cite{kang2023identity}, and the combination of secret sharing with homomorphic encryption to bolster data privacy~\cite{xie2022privacy}. Another approach involves using fully homomorphic encryption with multikey capabilities to secure data processing~\cite{liao2023multikey}, and optimizing federated learning to enhance privacy and efficiency in edge environments~\cite{he2022privacy}.}

\subsubsection{Data Privacy in Edge Computing}
In edge computing, ensuring data privacy is crucial due to decentralized processing and storage. Innovative approaches include partitioning deep neural networks to balance privacy with energy consumption~\cite{chaopeng2023privacy} and integrating federated learning with hybrid differential privacy and adaptive compression to protect data in industrial settings~\cite{jiang2021privacy}. Challenges in location-based services are addressed by crafting specific privacy protection strategies~\cite{liu2022privacy}, and a novel framework for secure data sharing emphasizes privacy-preserving computations across the computing landscape~\cite{zheng2023secure}. Additionally, Local Differential Privacy (LDP) techniques process data while preserving personal information, though they can reduce data utility due to noise addition~\cite{shin2018privacy}.

\subsubsection{Location Privacy in Edge Computing}
Ensuring location privacy within edge computing is paramount due to the proliferation of location-based services. Techniques include task offloading with reinforcement learning to protect privacy in autonomous transport systems~\cite{gao2022ppo2}, and dual K-anonymous caching strategies within edge environments to safeguard location data~\cite{zhang2023caching}. Dynamic strategies to protect location privacy in multi-user mobile edge systems highlight the need for agile response mechanisms in privacy protection~\cite{wang2023location}. Additional efforts focus on balancing service efficiency with privacy in location-based services~\cite{liu2022privacy} and task offloading frameworks that consider location privacy~\cite{wang2023location}. Vehicular crowdsourcing applications also reflect a focus on driver privacy, combining pseudonym and mix-zone techniques~\cite{ni2017privacy, basudan2017privacy}.

This literature review is summarized in Table~\ref{tab:privacy_edge_computing}.

Challenges: Despite its numerous advantages, edge computing faces several challenges, including security concerns, data privacy issues, and management complexities~\cite{jararweh2016future}. The distributed nature of edge computing introduces multiple points of vulnerability that must be secured against potential threats. Managing data privacy also becomes more complex as data are processed across numerous edge devices and locations. Solutions to these challenges include implementing advanced encryption techniques, establishing robust access control measures, and developing standardized protocols for data management and~security.

The increasing need to embed AI capabilities at the edge allows devices to perform sophisticated data analysis and decision-making processes locally. This enhances the autonomy of smart applications and reduces dependency on central servers. Furthermore, the ongoing development of 5G networks is expected to significantly bolster the capabilities of edge computing, enabling faster data processing and exchange rates.

Mobile crowdsourcing and edge computing have revolutionized data processing and service delivery by bringing computation closer to the data source. However, these technologies pose significant privacy challenges, particularly when handling sensitive user information such as personal identity and location data. This case study examines a smart city project that integrates edge computing and mobile crowdsourcing to improve public services like traffic management, waste disposal, and public safety. Residents contribute data via mobile applications, which are processed at distributed edge devices throughout the city. This initiative faces several privacy challenges, including identity privacy, data privacy, and location privacy.

\begin{table}[H]
 
 \caption{Privacy concerns and solutions in edge computing.}


\begin{tabularx}{\textwidth}{|Y|Y|Y|}
\toprule

\textbf{Privacy Concern} & \textbf{Research Focus} & \textbf{Challenges and Solutions} \\
\midrule
Identity Privacy& Pseudonym technology, cryptographic methods, data aggregation schemes~\cite{lysyanskaya2000pseudonym, wang2017secure, lu2011pseudonym} & 
\begin{itemize}\vspace{-6pt}
    \item De-anonymization attacks~\cite{gambs2014anonymization, narayanan2008robust, nilizadeh2014community}
    \item Limited data integrity~\cite{wang2017secure}
    \item Anonymous authentication using identity-based cryptography~\cite{kang2023identity}
    \item Secret sharing with homomorphic encryption~\cite{xie2022privacy}
    \item Fully homomorphic encryption with multikey capabilities~\cite{liao2023multikey}
    \item Optimized federated learning~\cite{he2022privacy} \vspace{-12pt}
\end{itemize} \\
\midrule
Data Privacy& Ensuring data privacy due to decentralized processing and storage & 
\begin{itemize}\vspace{-6pt}
    \item Partitioning deep neural networks for privacy and energy balance~\cite{chaopeng2023privacy}
    \item Federated learning with hybrid differential privacy and adaptive compression~\cite{jiang2021privacy}
    \item Location-based service privacy strategies~\cite{liu2022privacy}
    \item Secure data sharing frameworks~\cite{zheng2023secure}
    \item Local Differential Privacy (LDP) techniques~\cite{shin2018privacy}\vspace{-12pt}

\end{itemize} \\
\midrule 

Location Privacy& Protecting privacy in location-based services & 
\begin{itemize}\vspace{-6pt}
    \item Task offloading with reinforcement learning in autonomous transport systems~\cite{gao2022ppo2}
    \item Dual K-anonymous caching strategies~\cite{zhang2023caching}
    \item Dynamic multi-user privacy protection~\cite{wang2023location}
    \item Balancing service efficiency with location privacy~\cite{liu2022privacy}
    \item Task offloading frameworks for location privacy~\cite{wang2023location}
    \item Pseudonym and mix-zone techniques in vehicular crowdsourcing~\cite{ni2017privacy, basudan2017privacy}\vspace{-12pt}

\end{itemize} \\
\midrule
\end{tabularx}
\label{tab:privacy_edge_computing}
\end{table}

\section{Enhancing Privacy in Mobile Crowdsourcing and Edge Computing: A Case Study}
Mobile crowdsourcing and edge computing have transformed data processing and service delivery by making computation closer to the data source. However, these technologies also present privacy concerns, especially when handling sensitive user information like personal identity and location data. 
We consider a smart city project that integrates edge computing and mobile crowdsourcing to enhance public services such as traffic management, waste disposal, and public safety. Residents participate by sharing data via mobile applications, which are processed at edge devices distributed across the city. Traffic patterns are monitored using mobile data collected from citizens' smartphones, while waste management facilities use data from smart bins. As residents contribute sensitive information, this project faces the following privacy challenges: identity privacy (protecting citizens' identities), data privacy (secure processing of shared data), and location privacy (protecting individuals' exact whereabouts).

\subsection{Identity Privacy in Edge Computing} 
To protect identity privacy, the project implemented identity-based encryption \cite{yu2014identity} and forward security \cite{li2023privacy}. Forward security ensures that even if an encryption key is compromised, previously encrypted data remain secure. Additionally, decentralized identity management using blockchain-based~\cite{kumar2021blockchain} trust models was introduced to reduce reliance on centralized authorities. Each participant had a digital identity verified by decentralized nodes, mitigating risks of identity theft and unauthorized access.

\subsection{Data Privacy with Federated Learning}
For safeguarding data privacy, the project employed federated learning~\cite{zhao2022crowdfl} with hybrid differential privacy. ML models are trained on data collected from citizens without centralizing their data. Instead, model updates are shared from edge devices back to a central server. This reduces the risk of data exposure while still leveraging crowdsourced data for model improvements.

\subsection{Location Privacy with K-Anonymous Caching}
To protect location privacy, K-anonymous caching~\cite{zhang2023caching}was used to obfuscate user's exact locations by combining location data with that of other users in a given area. For instance, when a user reports parking availability, the system aggregates reports from multiple users in the vicinity to ensure individual anonymity.

\subsection{Recommendations}
The following recommendations are made to enhance the project's effectiveness:

\begin{itemize}
    \item {Establish clear guidelines for data collection, storage, and usage. Implement anonymization techniques and conduct regular audits to enhance transparency and accountability.}
    \item Integrate post-quantum cryptography~\cite{zanetti2022post} to future-proof encryption systems against quantum computing threats. Utilize privacy-preserving authentication (PPA)~\cite{liu2014shared} for efficient and secure data handling.
    \item Leverage blockchain for decentralized identity management to enhance trust models and improve data security with tamper-resistant record-keeping.
    \item Incorporate privacy considerations from the outset of projects through thorough privacy impact assessments and embrace principles like data minimization.
    \item Encourage public participation in decision-making processes related to data usage and enforce regulatory oversight to ensure compliance with evolving privacy standards.
    \item Focus on scalability by integrating AI and machine learning for dynamic risk detection in real time. Implement trusted execution environments for secure processing of sensitive data.
    \item Balance energy consumption with privacy needs using energy-efficient cryptographic methods that maintain security without compromising performance.
\end{itemize}

\section{Discussion}

This paper has explored the critical privacy concerns associated with mobile crowdsourcing and edge computing. As these technologies continue to evolve and become integral to modern data processing and storage solutions, addressing privacy concerns becomes paramount.

Identity privacy in cloud and edge computing is a multifaceted challenge. Research emphasizes robust forward security measures and identity-based encryption to protect sensitive user information~\cite{li2023privacy, veeresh2023identity, suresha2022ensuring}. Proposals such as CIDM aim to enhance security by separating authorization credentials and securing communication links~\cite{khalil2014consolidated}. Additionally, blockchain-based trust models and decentralized identity management systems offer promising approaches to reduce the need for trusted third parties, mitigating the risks of identity theft and unauthorized access~\cite{khajehei2018preserving, bendiab2018wip}. 
However, significant challenges remain, particularly in countering de-anonymization attacks and ensuring the integrity of anonymized data~\cite{gambs2014anonymization}. Ongoing research into advanced cryptographic methods and innovative authentication mechanisms, including those leveraging identity-based cryptography, is needed for addressing these vulnerabilities.

Ensuring data privacy in cloud and edge computing environments is critical due to the decentralized nature of these technologies. Innovative approaches such as partitioning deep neural networks to balance privacy with energy consumption and employing federated learning with hybrid differential privacy are making significant strides in this area~\cite{chaopeng2023privacy, jiang2021privacy}. Techniques like homomorphic encryption and LDP have shown potential in protecting data privacy by enabling secure data processing and analysis~\cite{wang2022dpp, shin2018privacy}. Despite their promise, these techniques often introduce challenges such as increased computational overhead and reduced data utility due to noise addition. 

The protection of location privacy is particularly crucial in mobile computing due to the pervasive use of location-based services. Research has proposed several methods to address this issue, including cipher-text retrieval techniques, spatial cloaking, and SHA-2 for robust mutual authentication~\cite{zhang2022location, jadallah2019spatial, rana2020mutual}. Strategies such as task offloading with reinforcement learning and dual K-anonymous caching have been explored to enhance location privacy in edge computing environments~\cite{gao2022ppo2, zhang2023caching}. These techniques offer dynamic and scalable solutions to protect location data while maintaining the efficiency of edge~services.

The implications of these privacy concerns are vast and multifaceted. As mobile computing becomes more ubiquitous, ensuring robust privacy protection mechanisms is not only a technical challenge but also a societal imperative. The interplay between user trust and technology adoption cannot be overlooked.  Compliance with the regulations such as  data protection laws requires the development of sophisticated privacy-preserving technologies that can adapt to diverse legal requirements and protect user data across different jurisdictions.

Future research should focus on enhancing the scalability and efficiency of privacy-preserving methods~\cite{wang2022bilateral,he2022privacy}, addressing the challenges of de-anonymization and data utility, and exploring new cryptographic techniques to ensure robust privacy protection. The integration of artificial intelligence and machine learning into privacy protection strategies offers promising avenues for innovation. These technologies can help identify and mitigate privacy risks in real time, providing an additional layer of security in dynamic and complex computing environments. Additionally, interdisciplinary collaboration among researchers, industry practitioners, and policymakers is essential to create standardized guidelines and frameworks for privacy protection in mobile computing. 

\subsection{Emerging Trends in Privacy Preservation} Some key trends and emerging technologies to enhance privacy while enabling valuable insights and innovations are the following: 

\begin{enumerate}
    \item Artificial Intelligence (AI) and ML are increasingly being used to enhance privacy protection. These methods and technologies are already being used for anomaly detection, breach prediction, and automation of privacy-preserving processes~\cite{yang2024ai}.
    \item Federated Learning is also a promising approach to privacy-preserving machine learning~\cite{kairouz2021advances}. This technique allows models to be trained on different distributed datasets without centralizing the data and addresses privacy concerns in sectors such as healthcare and finance. The process involves the following: (1) initial model training on a central server; (2) distribution of the model to individual devices; (3) local training on each device; (4) transmission of model updates back to the central server; (5) aggregation of updates to improve the global model.
    \item Encryption remains a main method of data protection, with many methods for enhancing its capabilities and applications. For example, homomorphic encryption~\cite{wang2022dpp} allows computations to be performed on encrypted data without decryption. This method allows secure data processing in untrusted environments, such as public~clouds.
    \item Post-Quantum Cryptography \cite{zanetti2022post} is used for encrypting algorithms that are resistant to attacks by both classical and quantum computers. Key developments include the following:
(1)~NIST standardization process for post-quantum cryptographic algorithms~\cite{nist_post_quantum}.  
(2) Integration of post-quantum algorithms into existing protocols~\cite{nist_post_quantum}. (3) Hybrid approaches combining classical and post-quantum methods.
    \item Privacy-Enhancing Computation (PEC)~\cite{nist_pec} covers a range of technologies that enable data to be processed or analyzed while keeping it encrypted or otherwise protected.
    \item Trusted Execution Environments (TEEs)~\cite{trustonic_tee} provide isolated execution environments within processors and allow sensitive computations to be performed securely even on untrusted systems. This allows hardware-level isolation of sensitive code and data and protection against privileged software attacks.
    \item Secure Multi-Party Computation (SMPC)~\cite{zhao2019secure} allows multiple parties to compute a function together over their inputs while keeping these inputs private. Some use cases include privacy-preserving benchmarking, secure auctions and voting systems, and machine learning in a collaborative environment across multiple data owners.
\end{enumerate}

\subsection{Interdisciplinary Efforts}
In the domains of mobile crowdsourcing and edge computing, fostering interdisciplinary collaboration is essential to create comprehensive frameworks that balance technological advancement with user privacy. Research is also directed towards decentralized systems, which enhance user control over data and mitigate risks associated with centralized data repositories by processing data closer to its source. Blockchain technology can be integrated for traceability and immutability in data handling practices and can enhance security and transparency, allowing users to verify how their data are used and shared. 

Incorporating energy efficiency into mobile crowdsourcing and edge computing is crucial for sustainable technology development~\cite{zhu2021green}. One effective strategy is the use of edge computing, which enhances energy efficiency by processing data closer to where it is generated. These methods can help in reducing the need for extensive data transmission and lowering overall energy consumption. This approach can help minimize latency and improve data processing speeds and can also significantly reduce the carbon footprint associated with centralized data centers.

\section{Conclusions}
This paper has examined the critical privacy concerns inherent in mobile crowdsourcing and edge computing, focusing on identity, data, and location privacy. Despite significant advancements, robust and scalable privacy-preserving methods are still required to address challenges such as de-anonymization attacks and data integrity issues. Ensuring user trust, complying with regulatory standards, and promoting the broader adoption of mobile computing technologies depend on effectively addressing these privacy concerns.
Future research should prioritize developing scalable and efficient privacy-preserving techniques and integrating artificial intelligence and machine learning for real-time privacy risk mitigation. Additionally, interdisciplinary collaboration among researchers, industry practitioners, and policymakers is essential to create standardized guidelines and frameworks for privacy protection.

 
\vspace{6pt}
\textbf{Funding: }{This research received no external funding}

\textbf{Conflict of Interest:}{The authors declare no conflict of interest}

\appendix

\begin{table}[h]
\footnotesize
\centering
\caption{Important Definitions}
\label{app:definitions}
\begin{tabular}{|p{2cm}|p{9cm}|}
\hline
\textbf{Keyword} & \textbf{Definition} \\
\hline
\textbf{Privacy} & The protection of sensitive information from unauthorized access, ensuring personal data such as identity, location, and activities are secure. \\
\hline
\textbf{Security} & Measures taken to protect systems and data from cyber-attacks, unauthorized access, and breaches in both mobile crowdsourcing and edge computing. \\
\hline
\textbf{Edge Computing} & A decentralized approach where data processing occurs closer to the data source (e.g., IoT devices) rather than relying on centralized cloud servers, enhancing performance and reducing latency. \\
\hline
\textbf{Mobile Crowdsourcing} & A method where data is collected from a large number of individuals through mobile devices, often for real-time applications such as traffic monitoring or environmental sensing. \\
\hline
\textbf{Location Privacy} & The protection of an individual's geographical location information to prevent unauthorized tracking or misuse. \\
\hline
\textbf{Data Privacy} & Ensuring that the collected data remains confidential and is not accessed or altered without permission, with special attention to sensitive information like health or financial data. \\
\hline
\textbf{Identity Privacy} & The safeguarding of personal identifiers such as names or social security numbers to prevent identity theft or unauthorized linking of activities to an individual. \\
\hline
\textbf{Cloud Computing} & The use of remote servers to store, manage, and process data, as opposed to local servers or personal devices, often integrated with edge computing for enhanced scalability. \\
\hline
\textbf{IoT (Internet of Things)} & The network of connected devices, such as smartphones and smart home appliances, that collect and exchange data through the internet. \\
\hline
\textbf{Differential Privacy} & A privacy technique that adds random noise to data sets, preventing the identification of individual data points while allowing analysis of aggregate trends. \\
\hline
\textbf{Pseudonymization} & The process of replacing private identifiers (like names) with pseudonyms to ensure anonymity while still allowing the data to be useful. \\
\hline
\textbf{Homomorphic Encryption} & A type of encryption that allows computations to be performed on encrypted data without needing to decrypt it, ensuring data privacy even during analysis. \\
\hline
\textbf{Federated Learning} & A machine learning technique that allows multiple devices to collaboratively train a model without sharing raw data, preserving data privacy. \\
\hline
\textbf{Sybil Attack} & A type of security attack where a single adversary creates multiple fake identities in a network to manipulate the system, commonly discussed in crowdsourcing security. \\
\hline
\textbf{Data Poisoning} & A malicious attack where false or misleading data is introduced into a dataset to corrupt the outcome of a crowdsourced or machine learning model. \\
\hline
\textbf{Spatial Cloaking} & A privacy technique used to blur or generalize location data to protect an individual’s exact location, commonly used in location-based services. \\
\hline
\textbf{Blockchain} & A decentralized digital ledger technology that securely records transactions across multiple systems, enhancing transparency and security, especially for data privacy. \\
\hline
\textbf{Encryption} & A method of converting readable data into an encoded format that can only be read by someone who has the correct decryption key, protecting data during storage and transmission. \\
\hline
\textbf{Task Offloading} & The process of transferring computational tasks from resource-limited devices (e.g., mobile devices) to more powerful edge servers, improving efficiency and performance. \\
\hline
\textbf{Trusted Execution Environment (TEE)} & A secure area of a device processor that ensures code and data loaded inside are protected from being tampered with the goal to provide a safe execution space for sensitive operations.\\
\hline
\end{tabular}

\end{table}
\bibliographystyle{abbrv}
\bibliography{REFERENCES}

\begin{thebibliography}{10}

\bibitem{waze}
Driving directions, live traffic \& road conditions updates - waze.
\newblock \url{https://www.waze.com/live-map/}, 2024.
\newblock Accessed on 01/20/2024.

\bibitem{a2020comprehensive}
Z.~A.~Almusaylim and N.~Jhanjhi.
\newblock Comprehensive review: Privacy protection of user in location-aware services of mobile cloud computing.
\newblock {\em Wireless Personal Communications}, 111:541--564, 2020.

\bibitem{abbas2017mobile}
N.~Abbas, Y.~Zhang, A.~Taherkordi, and T.~Skeie.
\newblock Mobile edge computing: A survey.
\newblock {\em IEEE Internet of Things Journal}, 5(1):450--465, 2017.

\bibitem{adomavicius2010context}
G.~Adomavicius and A.~Tuzhilin.
\newblock Context-aware recommender systems.
\newblock In {\em Recommender systems handbook}, pages 217--253. Springer, 2010.

\bibitem{ALI2023109605}
Y.~Ali and H.~U. Khan.
\newblock A survey on harnessing the applications of mobile computing in healthcare during the covid-19 pandemic: Challenges and solutions.
\newblock {\em Computer Networks}, 224:109605, 2023.

\bibitem{10169821}
M.~Barakat, R.~A. Saeed, and S.~Edam.
\newblock A comparative study on cloud and edgeb computing: A survey on current research activities and applications.
\newblock In {\em 2023 IEEE 3rd International Maghreb Meeting of the Conference on Sciences and Techniques of Automatic Control and Computer Engineering (MI-STA)}, pages 679--684, 2023.

\bibitem{bashir2022improving}
S.~R. Bashir and V.~B. Mi{\v{s}}i{\'c}.
\newblock Improving rating and relevance with point-of-interest recommender system.
\newblock In {\em ICC 2022-IEEE International Conference on Communications}, pages 1734--1739. IEEE, 2022.

\bibitem{bashir2023bert4loc}
S.~R. Bashir, S.~Raza, and V.~B. Misic.
\newblock Bert4loc: Bert for location—poi recommender system.
\newblock {\em Future Internet}, 15(6):213, 2023.

\bibitem{basudan2017privacy}
S.~Basudan, X.~Lin, and K.~Sankaranarayanan.
\newblock A privacy-preserving vehicular crowdsensing-based road surface condition monitoring system using fog computing.
\newblock {\em IEEE Internet of Things Journal}, 4(3):772--782, 2017.

\bibitem{bendiab2018wip}
K.~Bendiab, N.~Kolokotronis, S.~Shiaeles, and S.~Boucherkha.
\newblock Wip: A novel blockchain-based trust model for cloud identity management.
\newblock In {\em 2018 IEEE 16th Intl Conf on Dependable, Autonomic and Secure Computing, 16th Intl Conf on Pervasive Intelligence and Computing, 4th Intl Conf on Big Data Intelligence and Computing and Cyber Science and Technology Congress (DASC/PiCom/DataCom/CyberSciTech)}, pages 724--729. IEEE, 2018.

\bibitem{brakerski2011fully}
Z.~Brakerski and V.~Vaikuntanathan.
\newblock Fully homomorphic encryption from ring-lwe and security for key dependent messages.
\newblock In {\em Annual cryptology conference}, pages 505--524. Springer, 2011.

\bibitem{chaopeng2023privacy}
G.~Chaopeng, L.~Zhengqing, and S.~Jie.
\newblock A privacy protection approach in edge-computing based on maximized dnn partition strategy with energy saving.
\newblock {\em Journal of Cloud Computing}, 12(1):1--16, 2023.

\bibitem{cutillo2009privacy}
L.~A. Cutillo, R.~Molva, and T.~Strufe.
\newblock Privacy preserving social networking through decentralization.
\newblock In {\em 2009 Sixth International Conference on Wireless On-Demand Network Systems and Services}, pages 145--152. IEEE, 2009.

\bibitem{gambs2014anonymization}
S.~Gambs, M.-O. Killijian, and M.~N. del Prado~Cortez.
\newblock De-anonymization attack on geolocated data.
\newblock {\em Journal of Computer and System Sciences}, 80(8):1597--1614, 2014.

\bibitem{gao2022ppo2}
H.~Gao, W.~Huang, T.~Liu, Y.~Yin, and Y.~Li.
\newblock Ppo2: Location privacy-oriented task offloading to edge computing using reinforcement learning for intelligent autonomous transport systems.
\newblock {\em IEEE transactions on intelligent transportation systems}, 2022.

\bibitem{guan2018data}
Y.~Guan, J.~Shao, G.~Wei, and M.~Xie.
\newblock Data security and privacy in fog computing.
\newblock {\em IEEE Network}, 32(5):106--111, 2018.

\bibitem{guo2015mobile}
B.~Guo, Z.~Wang, Z.~Yu, Y.~Wang, N.~Y. Yen, R.~Huang, and X.~Zhou.
\newblock Mobile crowd sensing and computing: The review of an emerging human-powered sensing paradigm.
\newblock {\em ACM computing surveys (CSUR)}, 48(1):1--31, 2015.

\bibitem{he2022privacy}
C.~He, G.~Liu, S.~Guo, and Y.~Yang.
\newblock Privacy-preserving and low-latency federated learning in edge computing.
\newblock {\em IEEE Internet of Things Journal}, 9(20):20149--20159, 2022.

\bibitem{hu2010privacy}
L.~Hu and C.~Shahabi.
\newblock Privacy assurance in mobile sensing networks: Go beyond trusted servers.
\newblock In {\em 2010 8th IEEE International Conference on Pervasive Computing and Communications Workshops (PERCOM Workshops)}, pages 613--619. IEEE, 2010.

\bibitem{irshad2017advances}
S.~Irshad and D.~R.~A. Rambli.
\newblock Advances in mobile augmented reality from user experience perspective: a review of studies.
\newblock In {\em Advances in Visual Informatics: 5th International Visual Informatics Conference, IVIC 2017, Bangi, Malaysia, November 28--30, 2017, Proceedings 5}, pages 466--477. Springer, 2017.

\bibitem{jadallah2019spatial}
H.~Jadallah and Z.~Al~Aghbari.
\newblock Spatial cloaking for location-based queries in the cloud.
\newblock {\em Journal of Ambient Intelligence and Humanized Computing}, 10:3339--3347, 2019.

\bibitem{jararweh2016future}
Y.~Jararweh, A.~Doulat, O.~AlQudah, E.~Ahmed, M.~Al-Ayyoub, and E.~Benkhelifa.
\newblock The future of mobile cloud computing: integrating cloudlets and mobile edge computing.
\newblock In {\em 2016 23rd International conference on telecommunications (ICT)}, pages 1--5. IEEE, 2016.

\bibitem{jiang2021privacy}
B.~Jiang, J.~Li, H.~Wang, and H.~Song.
\newblock Privacy-preserving federated learning for industrial edge computing via hybrid differential privacy and adaptive compression.
\newblock {\em IEEE Transactions on Industrial Informatics}, 19(2):1136--1144, 2021.

\bibitem{kairouz2021advances}
P.~Kairouz, H.~B. McMahan, B.~Avent, A.~Bellet, M.~Bennis, A.~N. Bhagoji, K.~Bonawitz, Z.~Charles, G.~Cormode, R.~Cummings, et~al.
\newblock Advances and open problems in federated learning.
\newblock {\em Foundations and trends{\textregistered} in machine learning}, 14(1--2):1--210, 2021.

\bibitem{kang2023identity}
N.~Kang, Z.~Ning, S.~Zhang, M.~Waqas, et~al.
\newblock Identity-based edge computing anonymous authentication protocol.
\newblock {\em Tech Science Press}, 2023.

\bibitem{khajehei2018preserving}
K.~Khajehei.
\newblock Preserving privacy in cloud identity management systems using dcm (dual certificate management).
\newblock {\em Int. J. Wirel. Microw. Technol}, 8(4):54--65, 2018.

\bibitem{khalil2014consolidated}
I.~Khalil, A.~Khreishah, and M.~Azeem.
\newblock Consolidated identity management system for secure mobile cloud computing.
\newblock {\em Computer Networks}, 65:99--110, 2014.

\bibitem{ko2020lpga}
H.~Ko, H.~Lee, T.~Kim, and S.~Pack.
\newblock Lpga: Location privacy-guaranteed offloading algorithm in cache-enabled edge clouds.
\newblock {\em IEEE Transactions on Cloud Computing}, 10(4):2729--2738, 2020.

\bibitem{kong2019mobile}
X.~Kong, X.~Liu, B.~Jedari, M.~Li, L.~Wan, and F.~Xia.
\newblock Mobile crowdsourcing in smart cities: Technologies, applications, and future challenges.
\newblock {\em IEEE Internet of Things Journal}, 6(5):8095--8113, 2019.

\bibitem{kumar2021blockchain}
N.~Kumar, S.~Aggarwal, and P.~Raj.
\newblock {\em The blockchain technology for secure and smart applications across industry verticals}.
\newblock Academic Press, 2021.

\bibitem{li2023privacy}
F.~Li, J.~Wang, and Z.~Song.
\newblock Privacy protection of cloud computing based on strong forward security.
\newblock {\em International Journal of Cloud Applications and Computing (IJCAC)}, 13(1):1--9, 2023.

\bibitem{liao2023multikey}
J.~Liao, H.~Wang, and J.~Wu.
\newblock A multikey fully homomorphic encryption privacy protection protocol based on blockchain for edge computing system.
\newblock {\em Concurrency and Computation: Practice and Experience}, 35(4):e7539, 2023.

\bibitem{liu2014shared}
H.~Liu, H.~Ning, Q.~Xiong, and L.~T. Yang.
\newblock Shared authority based privacy-preserving authentication protocol in cloud computing.
\newblock {\em IEEE Transactions on parallel and distributed systems}, 26(1):241--251, 2014.

\bibitem{liu2021cloud}
Q.~Liu, J.~Gu, J.~Yang, Y.~Li, D.~Sha, M.~Xu, I.~Shams, M.~Yu, and C.~Yang.
\newblock Cloud, edge, and mobile computing for smart cities.
\newblock {\em Urban Informatics}, pages 757--795, 2021.

\bibitem{liu2022privacy}
S.~Liu.
\newblock Privacy protection and service evaluation methods for location-based services in edge computing environments.
\newblock {\em arXiv preprint arXiv:2212.03417}, 2022.

\bibitem{liu2019unitask}
Z.~Liu, Z.~Li, and K.~Wu.
\newblock Unitask: A unified task assignment design for mobile crowdsourcing-based urban sensing.
\newblock {\em IEEE Internet of Things Journal}, 6(4):6629--6641, 2019.

\bibitem{lu2011pseudonym}
R.~Lu, X.~Lin, T.~H. Luan, X.~Liang, and X.~Shen.
\newblock Pseudonym changing at social spots: An effective strategy for location privacy in vanets.
\newblock {\em IEEE transactions on vehicular technology}, 61(1):86--96, 2011.

\bibitem{10026418}
S.~Lu, J.~Lu, K.~An, X.~Wang, and Q.~He.
\newblock Edge computing on iot for machine signal processing and fault diagnosis: A review.
\newblock {\em IEEE Internet of Things Journal}, 10(13):11093--11116, 2023.

\bibitem{luo2021resource}
Q.~Luo, S.~Hu, C.~Li, G.~Li, and W.~Shi.
\newblock Resource scheduling in edge computing: A survey.
\newblock {\em IEEE Communications Surveys and Tutorials}, 23(4):2131--2165, 2021.

\bibitem{lysyanskaya2000pseudonym}
A.~Lysyanskaya, R.~L. Rivest, A.~Sahai, and S.~Wolf.
\newblock Pseudonym systems.
\newblock In {\em Selected Areas in Cryptography: 6th Annual International Workshop, SAC’99 Kingston, Ontario, Canada, August 9--10, 1999 Proceedings 6}, pages 184--199. Springer, 2000.

\bibitem{miao2023efficient}
Y.~Miao, Y.~Yang, X.~Li, L.~Wei, Z.~Liu, and R.~H. Deng.
\newblock Efficient privacy-preserving spatial data query in cloud computing.
\newblock {\em IEEE Transactions on Knowledge and Data Engineering}, 2023.

\bibitem{mukherjee2017security}
M.~Mukherjee, R.~Matam, L.~Shu, L.~Maglaras, M.~A. Ferrag, N.~Choudhury, and V.~Kumar.
\newblock Security and privacy in fog computing: Challenges.
\newblock {\em IEEE Access}, 5:19293--19304, 2017.

\bibitem{narayanan2008robust}
A.~Narayanan and V.~Shmatikov.
\newblock Robust de-anonymization of large sparse datasets.
\newblock In {\em 2008 IEEE Symposium on Security and Privacy (sp 2008)}, pages 111--125. IEEE, 2008.

\bibitem{ni2017privacy}
J.~Ni, K.~Zhang, X.~Lin, Q.~Xia, and X.~S. Shen.
\newblock Privacy-preserving mobile crowdsensing for located-based applications.
\newblock In {\em 2017 IEEE International Conference on Communications (ICC)}, pages 1--6. IEEE, 2017.

\bibitem{nilizadeh2014community}
S.~Nilizadeh, A.~Kapadia, and Y.-Y. Ahn.
\newblock Community-enhanced de-anonymization of online social networks.
\newblock In {\em Proceedings of the 2014 acm sigsac conference on computer and communications security}, pages 537--548, 2014.

\bibitem{nist_post_quantum}
N.~I. of~Standards and Technology.
\newblock Post-quantum cryptography.
\newblock \url{https://csrc.nist.gov/projects/post-quantum-cryptography}, 2024.
\newblock Accessed: 2024-09-30.

\bibitem{nist_pec}
N.~I. of~Standards and Technology.
\newblock Privacy-enhancing cryptography (pec).
\newblock \url{https://csrc.nist.gov/projects/pec}, 2024.
\newblock Accessed: 2024-09-30.

\bibitem{rana2020mutual}
K.~Rana, H.~Yadav, and C.~Agrawal.
\newblock Mutual authentication and location privacy using hecc and sha 2 in mobile cloud computing environment.
\newblock In {\em 2020 6th International Conference on Advanced Computing and Communication Systems (ICACCS)}, pages 362--369. IEEE, 2020.

\bibitem{raza2019progress}
S.~Raza and C.~Ding.
\newblock Progress in context-aware recommender systems—an overview.
\newblock {\em Computer Science Review}, 31:84--97, 2019.

\bibitem{ren2015exploiting}
J.~Ren, Y.~Zhang, K.~Zhang, and X.~Shen.
\newblock Exploiting mobile crowdsourcing for pervasive cloud services: challenges and solutions.
\newblock {\em IEEE Communications Magazine}, 53(3):98--105, 2015.

\bibitem{engproc2023032022}
B.~Sheikh, A.~Butt, and J.~Hanif.
\newblock Mobile cloud computing: A survey on current security trends and future directions.
\newblock {\em Engineering Proceedings}, 32(1), 2023.

\bibitem{shin2018privacy}
H.~Shin, S.~Kim, J.~Shin, and X.~Xiao.
\newblock Privacy enhanced matrix factorization for recommendation with local differential privacy.
\newblock {\em IEEE Transactions on Knowledge and Data Engineering}, 30(9):1770--1782, 2018.

\bibitem{EdgeAI2023}
R.~Singh and S.~S. Gill.
\newblock Edge ai: A survey.
\newblock {\em Internet of things and cyber-physical systems}, 2023.

\bibitem{suresha2022ensuring}
D.~Suresha, K.~Karibasappa, et~al.
\newblock Ensuring privacy preservation access control mechanism in cloud based on identity based derived key.
\newblock {\em International Journal of Advanced Computer Science and Applications}, 13(3), 2022.

\bibitem{trustonic_tee}
Trustonic.
\newblock What is a trusted execution environment (tee)?
\newblock \url{https://www.trustonic.com/technical-articles/what-is-a-trusted-execution-environment-tee/}, 2024.
\newblock Accessed: 2024-09-30.

\bibitem{veeresh2023identity}
V.~Veeresh and L.~R. Parvathy.
\newblock Identity-based encryption to implement anti-collusion information sharing schemes in cloud computing.
\newblock In {\em 2023 2nd International Conference on Applied Artificial Intelligence and Computing (ICAAIC)}, pages 1177--1182. IEEE, 2023.

\bibitem{wang2022bilateral}
H.~Wang, Y.~Yang, E.~Wang, X.~Liu, J.~Wei, and J.~Wu.
\newblock Bilateral privacy-preserving worker selection in spatial crowdsourcing.
\newblock {\em IEEE Transactions on Dependable and Secure Computing}, 2022.

\bibitem{wang2022dpp}
J.~Wang, F.~Wu, T.~Zhang, and X.~Wu.
\newblock Dpp: Data privacy-preserving for cloud computing based on homomorphic encryption.
\newblock In {\em 2022 International Conference on Cyber-Enabled Distributed Computing and Knowledge Discovery (CyberC)}, pages 29--32. IEEE, 2022.

\bibitem{wang2017secure}
L.~Wang, G.~Liu, and L.~Sun.
\newblock A secure and privacy-preserving navigation scheme using spatial crowdsourcing in fog-based vanets.
\newblock {\em Sensors}, 17(4):668, 2017.

\bibitem{wang2023location}
W.~Wang, X.~Zhou, T.~Qiu, X.~He, and S.~Ge.
\newblock Location privacy-aware service migration against inference attacks in multi-user mec systems.
\newblock {\em IEEE Internet of Things Journal}, 2023.

\bibitem{xie2022privacy}
H.~Xie, Y.~Guo, H.~Wang, Q.~Chen, C.~Fang, and N.~Zhu.
\newblock Privacy-preserving method of edge computing based on secret sharing and homomorphic encryption.
\newblock In {\em International Conference on Cloud Computing, Internet of Things, and Computer Applications (CICA 2022)}, volume 12303, pages 79--85. SPIE, 2022.

\bibitem{yang2024ai}
L.~Yang, M.~Tian, D.~Xin, Q.~Cheng, and J.~Zheng.
\newblock Ai-driven anonymization: Protecting personal data privacy while leveraging machine learning.
\newblock {\em arXiv preprint arXiv:2402.17191}, 2024.

\bibitem{yousefpour2019all}
A.~Yousefpour, C.~Fung, T.~Nguyen, K.~Kadiyala, F.~Jalali, A.~Niakanlahiji, J.~Kong, and J.~P. Jue.
\newblock All one needs to know about fog computing and related edge computing paradigms: A complete survey.
\newblock {\em Journal of Systems Architecture}, 98:289--330, 2019.

\bibitem{yu2014identity}
Y.~Yu, Y.~Mu, J.~Ni, J.~Deng, and K.~Huang.
\newblock Identity privacy-preserving public auditing with dynamic group for secure mobile cloud storage.
\newblock In {\em Network and System Security: 8th International Conference, NSS 2014, Xi’an, China, October 15-17, 2014, Proceedings 8}, pages 28--40. Springer, 2014.

\bibitem{zanetti2022post}
M.~Zanetti and I.~W{\"u}rth.
\newblock {\em Post Quantum Cryptography}.
\newblock PhD thesis, OST Ostschweizer Fachhochschule, 2022.

\bibitem{zeyu2020survey}
H.~Zeyu, X.~Geming, W.~Zhaohang, and Y.~Sen.
\newblock Survey on edge computing security.
\newblock In {\em 2020 International Conference on Big Data, Artificial Intelligence and Internet of Things Engineering (ICBAIE)}, pages 96--105. IEEE, 2020.

\bibitem{zhang2018data}
J.~Zhang, B.~Chen, Y.~Zhao, X.~Cheng, and F.~Hu.
\newblock Data security and privacy-preserving in edge computing paradigm: Survey and open issues.
\newblock {\em IEEE access}, 6:18209--18237, 2018.

\bibitem{zhang2023caching}
S.~Zhang, B.~Hu, W.~Liang, K.-C. Li, and B.~B. Gupta.
\newblock A caching-based dual k-anonymous location privacy-preserving scheme for edge computing.
\newblock {\em IEEE Internet of Things Journal}, 10(11):9768--9781, 2023.

\bibitem{zhang2022location}
Y.~Zhang, Q.~Zhang, Y.~Jiang, and Y.~Yan.
\newblock A location privacy protection method based on cipher-text retrieval in cloud environment.
\newblock {\em Security and Privacy}, 5(5):e250, 2022.

\bibitem{zhao2022crowdfl}
B.~Zhao, X.~Liu, W.-N. Chen, and R.~Deng.
\newblock Crowdfl: privacy-preserving mobile crowdsensing system via federated learning.
\newblock {\em IEEE Transactions on Mobile Computing}, 2022.

\bibitem{zhao2019secure}
C.~Zhao, S.~Zhao, M.~Zhao, Z.~Chen, C.-Z. Gao, H.~Li, and Y.-a. Tan.
\newblock Secure multi-party computation: theory, practice and applications.
\newblock {\em Information Sciences}, 476:357--372, 2019.

\bibitem{zheng2023secure}
K.~Zheng, C.~Ding, and J.~Wang.
\newblock A secure data-sharing scheme for privacy-preserving supporting node--edge--cloud collaborative computation.
\newblock {\em Electronics}, 12(12):2737, 2023.

\bibitem{zhu2021green}
S.~Zhu, K.~Ota, and M.~Dong.
\newblock Green ai for iiot: Energy efficient intelligent edge computing for industrial internet of things.
\newblock {\em IEEE Transactions on Green Communications and Networking}, 6(1):79--88, 2021.

\end{thebibliography}

\end{document}